\documentclass[prl,aps,twocolumn]{revtex4}
\usepackage{graphicx}

\begin{document}


\title {Thermal photon-IMF anticorrelation: a signal of prompt multifragmentation?}

\author { R.~Alba,$^1$ C.~Agodi,$^1$ C.~Maiolino,$^1$ A.~Del~Zoppo,$^1$
M.~Colonna,$^1$
G.~Bellia,$^{1,2}$ M.~Bruno,$^3$ N.~Colonna,$^4$
R. Coniglione,$^1$ M.~D'Agostino,$^3$
M.L.~Fiandri,$^3$ P.~Finocchiaro,$^1$ F.~Gramegna,$^5$
I.~Iori,$^6$ K.~Loukachine,$^1$
G.V.~Margagliotti,$^7$ P.F.~Mastinu,$^5$ P.M.~Milazzo,$^7$ E.~Migneco,$^{1,2}$
A.~Moroni,$^6$
P.~Piattelli,$^1$ R.~Rui,$^7$ D.~Santonocito,$^1$ P. Sapienza,$^1$
G.~Vannini$^3$ }

\address { (1) INFN - Laboratori Nazionali del Sud,
            Via S. Sofia 62, I-95123 Catania (ITALY)}
\address { (2) Dipartimento di Fisica e Astrnomia dell'Universit\`a, Catania (ITALY)}
\address { (3) INFN and Dipartimento di Fisica dell'Universit\`a, Bologna (ITALY)}
\address { (4) INFN, Bari (ITALY)}
\address { (5) INFN - Laboratori Nazionali di Legnaro, Legnaro (ITALY)}
\address { (6) INFN and Dipartimento di Fisica dell'Universit\`a, Milano (ITALY)}
\address { (7) INFN and Dipartimento di Fisica dell'Universit\`a, Trieste (ITALY)}

\date{\today}


\begin{abstract}
The mechanism responsible for
IMF emission in central $^{58}$Ni + $^{197}$Au reactions at
30 and 45 MeV/nucleon is investigated by looking at the thermal bremsstrahlung photon
production.
An IMF - photon anticorrelation signal is
observed, for central collisions, at 45 MeV/nucleon
with
IMF velocity around the center of mass value.
This observation is proposed as an
evidence for prompt nuclear fragmentation events.

\end{abstract}

\maketitle

Heavy ion collisions  at
intermediate energy offer the possibility to access states of
nuclear matter in  which density and temperature conditions, far from
normal ones, give rise to
a variety of new phenomena.

For instance, when the nuclear matter is compressed, two body
nucleon-nucleon collisions are favored and this leads  to an
enhanced production of incoherent bremsstrahlung photons. In the
Fermi energy domain, two main sources of incoherent bremsstrahlung
photons ($E_{\gamma}>25$ MeV) have been observed. In the first one,
located in the early compression phase of the reaction and in the
overlap region of the two colliding nuclei
\cite{Nif1,Mig,Mart,Sap,Nif2,Cass}
high energy 
bremsstrahlung photons are produced
in first chance collisions of projectile protons
(neutrons) on target neutrons (protons)  \cite{Sap,bauer,biro}. These are called
{\it direct} photons.
In the second one, that as indicated by interferometry measurements  \cite{Marq1}
has a larger extension in space-time with respect to the direct one \cite{Marq},
 bremsstrahlung photons are still produced in
p-n collisions but in a phase of the reaction in which the
identity of  projectile (target) nucleons is lost.
This results in a smaller photon energy, that reflects the temperature evolution of the composite system \cite{Schub,Mart2}. These are called {\it thermal} photons. 
A possible
interpretation of this phenomenon, substanciated by quantitative estimates \cite{Mart2,Marq},
is provided by  BUU calculations \cite{bona}. These studies associate thermal photon emission  with monopole oscillations (a second
compression phase) experienced by the composite system. This scenario appears quite appealing
since it opens the possibility to extract information on the nuclear matter compressibility and
equation of state.

Reactions at intermediate energies are also accompanied by a copious
production of IMF's (fragments with charge $3\le Z\le 20$), the so-called
multifragmentation \cite{Bowm}.
Whether such a process is a phenomenon consisting in the  prompt
disintegration of the system
during the expansion phase that follows the initial collisional shock
\cite{Aiche,Bondo,Gross,Hahn,Colo,Fried,Ono}
or it occurs at larger times, as a late deexcitation of
a hot thermalized system \cite{seq}, is still an open problem.

In this work we investigate experimentally, in heavy ion central collisions, the production of  IMF's and thermal photons inside the same reaction event.
The observable adopted in our study
is the
thermal photon-IMF multiplicity correlation factor
\begin{equation}
(1+R)_{th}=
{ { <m_{\gamma}^{th}\cdot m_{IMF}> }\over { <m_{\gamma}^{th}>\cdot <m_{IMF}>}}
\label{eq1}
\end{equation}
$m_{\gamma}^{th}$ being mostly 0 or 1.

Lack of correlation indicates unambigously that IMF and thermal photon production mechanisms are
compatible and independent. This is the case of  IMF emission in
the late de-excitation of a hot heavy system,
indicated as HHS in the following,
that is formed during the reaction path, and that can
be identified as a composite source in central collisions or as a target remnant in the other cases.
On the other hand,
if anticorrelation is observed, IMF and thermal photon emissions are incompatible mechanisms, like in a scenario where the prompt disintegration of the system, expected  in
violent collisions \cite{Aiche,Bondo,Gross,Hahn,Colo,Fried,Ono},
inhibits the formation
of  the thermal photon source, i.e. the HHS.
The value of the factor (1) equals zero
in the case of  prompt complete
disintegration of the system  into
small fragments (production of the coincident thermal photon totally inhibited),
while it equals 1 in the case of late statistical fragmentation of the
HHS, if the
energy balance does not introduce strong constraints.

We have studied the reaction $^{58}$Ni + $^{197}$Au
at two bombarding energies, 30 and 45 MeV/nucleon.
This choice
has been motivated by the need of a compromise between
the possibility to reach enough compression in the collision and a significant
thermal to direct photon ratio, that has been observed to decrease
with increasing
bombarding energy \cite{Mart2}.
Moreover, as we will see,
this study performed at two bombarding energies, allows
to rule out that the anticorrelation signal  is a trivial energy
balance effect which accounts for the energy taken by the photon.

The experiment has been performed at Laboratori Nazionali del Sud bombarding
a 2.5 mg/cm$^2$ thick Au target with
$^{58}$Ni beams at 30 and 45 MeV/nucleon delivered by the Tandem and
Superconductive
Cyclotron (CS) acceleration system.
Since nuclear bremsstrahlung is an extremely rare process, this study has
required the combined use of MEDEA \cite{Mig2} and MULTICS \cite{Iori}
sophisticated and extensive multidetector arrays.
MEDEA is a ball, built with 180 BaF$_2$, each 20 cm thick, placed at
22 cm from the target, which covers the polar angles from 30$^o$ to 170$^o$.
The BaF$_2$ permit to detect and identify light charged particles
and photons.
The MULTICS array consists of 55 telescopes, each formed by an
ionization chamber, a silicon detector and a CsI crystal, located at 50 cm
from the target, and allows the
identification of charged particles up to Z=83.
The total geometric acceptance is larger than 90$\%$ of 4$\pi$.

To select
the events as a function of centrality  we have used the
charged particle multiplicity, converted into impact parameter
according to the geometrical prescriptions of \cite{Cava}.
In the following analysis we will consider, as central events, the
impact parameter region $b/b_{max} < 0.35$. 
This corresponds to about $10^7$ events   
at the larger beam energy (45 MeV/nucleon). 
In this class of events the average IMF multiplicity is equal to $0.623 
\pm 4~ 10^{-3}$. 

\begin{figure}
\centering
\includegraphics[scale=0.5]{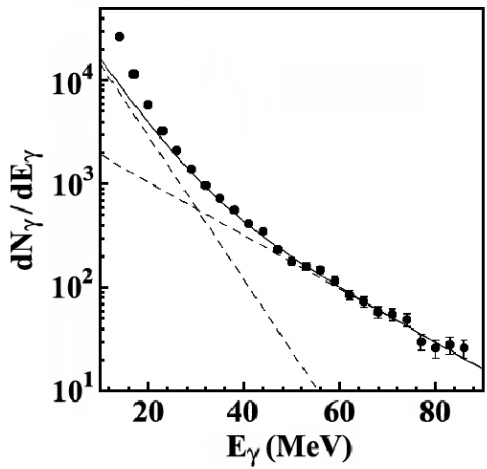}
\caption{\it Experimental photon energy spectrum for Ni + Au at 45 MeV/nucleon,
central reactions. Dashed lines: thermal and
direct contributions to the fit (full line).}
\label{spettri}
\end{figure}
Fig.1 shows the experimental
 photon energy spectrum
for the system $^{58}$Ni + $^{197}$Au at 45 MeV/nucleon,  obtained in
central events.
The photon spectrum deviates from a pure exponential dependence on
$E_{\gamma}$. Such a deviation is consistent with the presence of two
contributions, each one having a
dependence of the type $\exp { -{ {E_{\gamma}}\over {E_0} } }$ on $E_{\gamma}$
but with very different inverse slope parameters $E_0$ \cite{Alba}.
The two
component fit to the data is also shown in Fig.1.
The component with the largest $E_0$ can be related to the direct photons
produced in first chance two-body
collisions between nucleons coming from the initially
separated projectile and target Fermi clouds in macroscopic relative motion. In this case,
for a given projectile-target combination,
$E_0$ only depends on the incident energy.
The small $E_0$ component is associated with the thermal photons, the experimental values of
$E_0$ being related to the temperature
of the HHS \cite{taps_last,taps_new}.
At low energies ($E_{\gamma} <$  20 MeV) one observes also
statistical photons emitted
from the excited products. 

For both incident energies and selected impact
parameter $b$ bins,
a simultaneous fit to the photon energy and angular
distributions  allowed us to determine the relevant
characteristics of the thermal photon source \cite{Alba}. Inverse slope
parameter values of
the order of 4-5 MeV,
a source velocity approaching that of the nucleus-nucleus center of mass with
increasing centrality and a very small anisotropy have been deduced in
agreement with \cite{Schub,Mart2}. Thermal photons are present in all
the spectra of this experiment
indicating
that in the studied reactions the HHS
is formed in a significant number of events.

It is evident from Fig.1 that thermal photons cannot be isolated completely
from direct photons or statistical photons
and therefore the experimental correlation factor is the sum of the
correlation factors of the IMF's
with thermal, direct and statistical photons, weighted by the relative
intensities:
$$ 
(1+R)_{exp}={{I_{\gamma}^{th}}\over {I_{\gamma}}} \cdot(1+R)_{th}+
{{I_{\gamma}^{dir}}\over {I_{\gamma}}} \cdot(1+R)_{dir}+
$$
\begin{equation}
{{I_{\gamma}^{stat}}\over {I_{\gamma}}} \cdot(1+R)_{stat}.
\label{eq2}
\end{equation}

Let us focus on central collisions, in which about 8500 and 3000 $\gamma$'s with energy larger 
than 30 MeV have been collected at the incident energy of 45 MeV/nucleon and 30 MeV/nucleon,
respectively. 
The correlation factor (2) has been evaluated as a function of the
photon energy. Namely we consider
all photons with energy greater than a threshold
$E_\gamma$ and calculate the correlation factor as a function of $E_\gamma$.
We have considered two windows for the
IMF's velocity:
$W_{cm}$, that extends  up to $\beta \approx 0.1$ and includes
the nucleus-nucleus center of mass velocity,
i.e. the window where the thermal photon source is localized,
and
$W_{h}$, that includes all IMF's with higher velocity ($\beta > 0.1$).

The results obtained at 30 and 45 MeV/nucleon are displayed in Fig.2.
\begin{figure}
\centering
\includegraphics[scale=1.1]{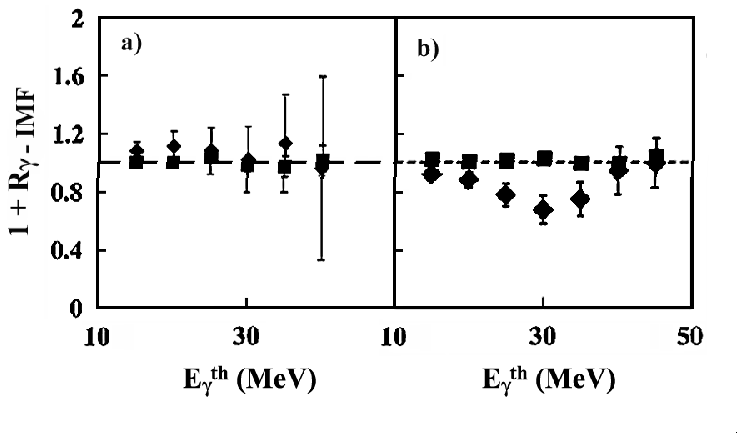}
\caption{\it Experimental photon - IMF correlation factors versus
the threshold $E_{\gamma}$ (see text), for IMF's
in two velocity windows: $W_{cm}$ (diamonds), $W_h$ (squares).
Data are shown at 30 MeV/nucleon (a)
and 45 MeV/nucleon (b), for central collisions.}
\label{corr_cent}
\end{figure}

At 45 MeV/nucleon, for IMF's in the $W_{cm}$ window, a
clear anticorrelation signal appears in the region where thermal
photon contribution is more abundant. 
This observation 
is the type of signal strongly expected in presence of
prompt fragmentation, that competes with the HHS formation and
inhibits thermal photon production. 
The signal disappears at low
and high gamma energy  due to the overbalance of the other photon
production mechanisms (see Eq.(2)), that also dims the true
anticorrelation signal. The IMF's with a larger velocity (in the $W_h$
window), mostly coming from pre-equilibrium emission  in the very
early stage of the collision, do not appear correlated with the
thermal photon production, as expected, because they are associated with a
different emission source.  At 30
MeV/nucleon, within error bars, $1 + R$ is always equal to
one, both in the window $W_h$ 
and in the window $W_{cm}$.
In this second case, the absence  of the
correlation can be seen as a signal of independent
production of gamma's and IMF's, due to  
the dominance of late fragmentation of the HHS. 

We underline that the difference observed at the two beam energies
cannot be ascribed to a trivial effect of the energy conservation,
because such effect works in the opposite direction and, due to the
energy carried away by the photon, should result in an
anticorrelation signal stronger at 30 MeV/nucleon than at 45
MeV/nucleon. 

At 45 MeV/nucleon
the analysis has been performed also as a function of the
impact parameter for a $\gamma$ energy bin in which the thermal
photon contribution is large (25 MeV $<E_\gamma<$ 40 MeV) and for IMF's in
$W_{cm}$. The results are shown in Fig.3. It can be
seen that the anticorrelation  increases with the violence of the
collision showing that the contribution of IMF's promptly emitted
grows with the centrality of the collision. The same analysis could
not be done at 30 MeV/nucleon for lack of statistics.

A rough evaluation of the significance of the differences 
observed in Fig.2 between the results at 30 and 45 MeV/nucleon
has been made according to the Kolmogorov test. The two trends have been found different at a confidence level of 95$\% $, in spite of the quite large error bars affecting the 30 MeV/nucleon data.

We have estimated the probability that the observed anticorrelation could be due to
multiple firing of the detectors, that could cause the lack of $\gamma$'s when
high LCP multiplicities (as in central collisions) are detected in the $BaF_2$ arrays.
For the detected multiplicities ($<20$) and the MEDEA granularity,  this probability results negligeable and affects the correlation signal by less than $3\%$.


\begin{figure}
\centering
\includegraphics[scale=1.7]{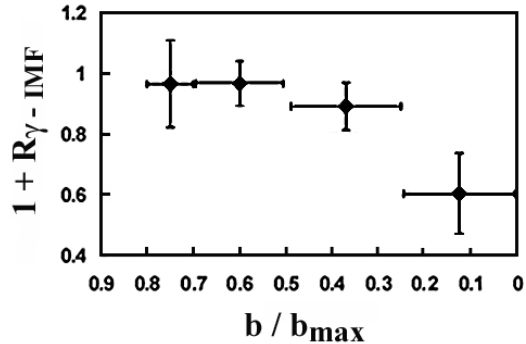}
\caption
{\it Thermal photon - IMF correlation factors, for IMF's
in the velocity window $W_{cm}$, as a function of
impact parameter. Data are shown at 45 MeV/nucleon.}
\label{corr_b}
\end{figure}

Prompt fragmentation processes have been recently
studied in the context of dynamical models incorporating the effects of
many-body correlations and fluctuations \cite{Colo,Ono}.
To check the interpretation proposed above,
stochastic mean field simulations \cite{Colo} have been performed
in the case of the reactions studied here at both incident energies, for
central impact parameters ( $b/b_{max} = 0.17$ in the calculations).
A soft EOS, with compressibility modulus $K = 200 MeV$, has been used.
A stiffer behaviour of the EOS would move the onset of prompt multifragmentation to
higher beam energies.
\begin{figure}
\centering
\includegraphics[scale=1.4]{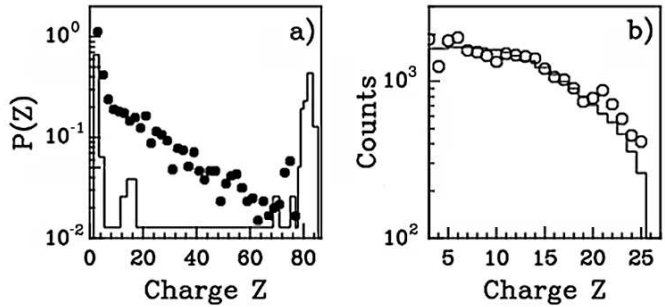}
\caption{\it
a) Primary fragment Z-distributions from stochastic mean-field
calculations at 30 MeV/nucleon (full line) and 45 MeV/nucleon (filled circles).
b) Comparison between experimental charge distribution, at 45 MeV/A (open circles),
and the final filtered simulations (full line). See text for details.}
\label{distri_new}
\end{figure}

It appears that
the primary fragment charge distributions predicted by these
calculations are completely different for the two energies considered (Fig.4a),
indicating the dominant role of prompt multifragmentation
only at 45 MeV/nucleon.
In fact, at 30 MeV/nucleon in most of the events, after the initial collisional shock
and the subsequent expansion, the system compacts again leading to
the formation of a HHS within a narrow interval of
charge  around $Z=80$.
On the other hand, at 45 MeV/nucleon this happens only in
a small fraction of the total number of events, leading to HHS's with charge
$Z\ge70$, while prompt IMF formation is mostly observed.
The final charge distribution, after evaporation of the primary products  \cite{SIMO}, has been calculated for both energies. At 30 MeV/nucleon IMF's are mainly produced by the
statistical decay of the HHS. 
Instead, at 45 MeV/nucleon the charge distribution is almost not affected by the secondary decay.
 In Fig.4b we show the final filtered distribution compared to the experimental data.
The two distributions have been normalized to the same area in the region $4<Z<25$.
The agreement is quite satisfactory. 
Moreover, the calculated filtered
IMF's multiplicities reproduce the experimental ones within $30\%$, a good agreement if we consider that the calculations are not impact parameter averaged.

Since, as discussed above, prompt IMF's and HHS are expected to be anticorrelated as IMF's and
thermal photons,
we have calculated the multiplicity correlation
factor
\begin{equation}
(1+R)_{HHS}=
{ { <m_{HHS}\cdot m_{IMF}> }\over { <m_{HHS}>\cdot <m_{IMF}>}},
\end{equation}
the possible values of $m_{HHS}$ being only 0 or 1.
As intuitively expected looking at the distributions
in Fig.4a, the anticorrelation results weak at 30 MeV/nucleon
($(1+R)_{HHS} {= 0.97} $) and
becomes stronger at 45 MeV/nucleon ($(1+R)_{HHS} {= 0.5}$).  

Hence simulations follow the same trend as observed in the experimental
data, supporting the occurrence of prompt multifragmentation at the highest
energy.
The prompt character of IMF emission has been also assessed in other recent
experimental observations \cite{Borderie,Michela}.

In summary, in this work we investigate the correlation factor between
thermal photons and IMF's as an indicator of the mechanism and of the time scales of 
nuclear fragmentation. For central Ni+Au collisions
the observation of an anticorrelation with IMF's in the center of mass
velocity window
appears as a
signal of prompt multifragmentation
at 45 MeV/nucleon.
Dynamical calculations based on the stochastic mean-field approach, with a soft EOS,
including statistical de-excitation of the primary HHS,
also indicate the dominance of prompt IMF emission in central collisions at 45 MeV/nucleon.


In conclusion, we believe that at the highest bombarding energy studied
in this experiment we have observed
different processes which mark significantly the dynamical evolution
of  violent nucleus-nucleus collisions: 
the production of direct photons in the primordial compressed phase, the onset of
prompt fragmentation during the expansion of the system and the competition with
a resilience of the nuclear system towards a composite source formation,
responsible for
the production of thermal photons.

We thank the LNS staff for the beam quality and for the support during
the experiment.






\end {document}